\documentclass[10pt,final,journal,letterpaper,oneside,onecolumn]{IEEEtran}
\usepackage[pdftex]{graphicx}
\pdfoutput=1
\usepackage[fleqn]{amsmath}
\usepackage[psamsfonts]{amssymb}

\def\x{\boldsymbol{x}}
\def\y{\boldsymbol{y}}
\def\z{\boldsymbol{z}}

\def\X{\boldsymbol{X}}

\def\0{\boldsymbol{0}}

\def\cB{\mathcal B}

\def\cI{\mathcal I}

\def\cP{\mathcal P}

\def\cS{\mathcal S}
\def\cT{\mathcal T}

\def\cV{\mathcal V}

\def\cX{\mathcal X}
\def\cY{\mathcal Y}

\def\bcS{\boldsymbol{\cS}}
\def\ov{\overline}
\def\ul{\underline}
\def\wh{\widehat}
\def\wt{\widetilde}

\def\eqdef{{\displaystyle\mathop{=}^{\mbox{\rm def.}}}}

\newcommand\dotdot[1]{\mbox{$\ddot{\mbox{#1}}$}}
\newcommand\dash[1]{\mbox{$\acute{\mbox{#1}}$}}%
\newtheorem{teiri}{Theorem}
\newtheorem{kei}{Corollary}
\newtheorem{hodai}{Lemma}
\newtheorem{teigi}{Definition}
\newtheorem{hosoku}{Remark}
\begin{document}
\allowdisplaybreaks  % allow to repage in mathematical environments

\title{%
  Universal source coding
  over generalized complementary delivery networks
}%
\author{%
  Akisato~Kimura,~\IEEEmembership{Senior Member,~IEEE},
  Tomohiko~Uyematsu,~\IEEEmembership{Senior Member,~IEEE},
  Shigeaki~Kuzuoka,~\IEEEmembership{Member,~IEEE}, and
  Shun~Watanabe,%~\IEEEmembership{Member,~IEEE}
  \thanks{%
    A. Kimura is with NTT Communication Science Laboratories,
    NTT Corporation, 3-1 Morinosato Wakamiya, Atsugi-shi, Kanagawa, 243-0198 Japan.
    E-mail: research@akisato.org
  }% <-this % stops a space
  \thanks{%
    T. Uyematsu and S. Watanabe are with Department of Communications and Integrated
    Systems, Tokyo Institute of Technology, 2-12-1 Ookayama, Meguro-ku, Tokyo, 152-8552
    Japan.
    E-mail: uyematsu@ieee.org, shun-wata@it.ss.titech.ac.jp
  }% <-this % stops a space
  \thanks{%
    S. Kuzuoka is with Department of Computer and Communication Sciences, Wakayama
    University, 930 Sakaedani, Wakayama, Wakayama 640-8510 Japan.
    E-mail: kuzuoka@sys.wakayama-u.ac.jp
  }% <-this % stops a space
  \thanks{%
    Manuscript received October 26, 2007.
  }% <-this % stops a space
}% <-this % stops a space
\markboth{IEEE TRANSACTIONS ON INFORMATION THEORY,~Vol.~xxx, No.~xxx,~xxxxx~200x}%
{Kimura \MakeLowercase{\textit{et al.}}: Universal coding over complementary delivery networks}%
\maketitle
\begin{abstract}
This paper deals with a universal coding problem for a certain kind of multiterminal
source coding network called a generalized complementary delivery network. In this
network, messages from multiple correlated sources are jointly encoded, and each decoder
has access to some of the messages to enable it to reproduce the other messages.
Both fixed-to-fixed length and fixed-to-variable length lossless coding schemes are
considered. Explicit constructions of universal codes and the bounds of the error
probabilities are clarified by using methods of types and graph-theoretical analysis.
\end{abstract}

% For peer-review papers, you can put extra information on the cover
% For peer-review papers, inserts a page break and creates the second title.
% Will be ignored for other modes.
%\IEEEpeerreviewmaketitle

%%%%%%%%
\section{Introduction}

A coding problem for correlated information sources was first described and investigated
by Slepian and Wolf \cite{SlepianWolf}, and later, various coding problems derived from
that work were considered (e.g. Wyner \cite{SideInformationCoding:Wyner}, K\dotdot{o}rner
and Marton \cite{KornerMarton}, Sgarro \cite{SgarroCoding}). Meanwhile, the problem of
universal coding for these systems was first investigated by Csisz\dash{a}r and
K\dotdot{o}rner \cite{UniversalSlepianWolf:Csiszar}. Universal coding problems are not
only interesting in their own right but are also very important in terms of practical
applications. Subsequent work has mainly focused on the Slepian-Wolf network
\cite{UniversalSlepianWolf:Csiszar2,UniversalSlepianWolf:Oohama,%
UniversalSlepianWolf:Uyematsu} since it appears to be difficult to construct universal
codes for most of the other networks. For example, Muramatsu \cite{PhD:muramatsu}
showed that no fixed-to-variable length (FV) universal code can attain the optimal
coding rate for the Wyner-Ziv coding problem\cite{WynerZiv}.

\begin{figure}[t]
  \begin{center}
    \includegraphics[width=0.5\hsize]{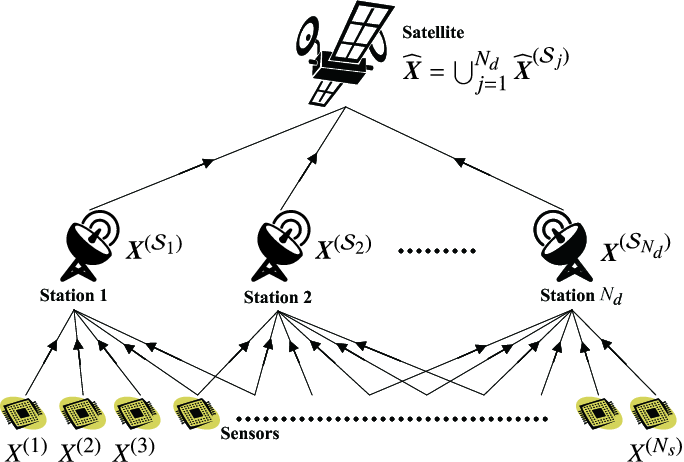}
    \caption{
      Data collection:
      Stations are physically separated from each other. Each station collects its own
      target data, and transmits them to a satellite.
    }
    \label{fig:satellite:1}\vspace{5mm}
    \includegraphics[width=0.5\hsize]{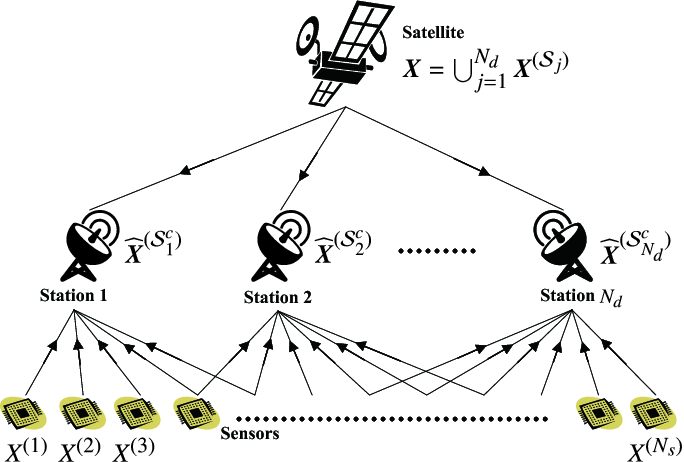}
    \caption{
      Data distribution:
      The satellite broadcasts the collected data back to the stations for sharing. Each
      station has already gathered its own target data, and thus wants to reproduce the
      other data by using its own target data as side information.
    }
    \label{fig:satellite:2}
  \end{center}
\end{figure}

\begin{figure}[t]
  \begin{center}
    \includegraphics[width=0.5\hsize]{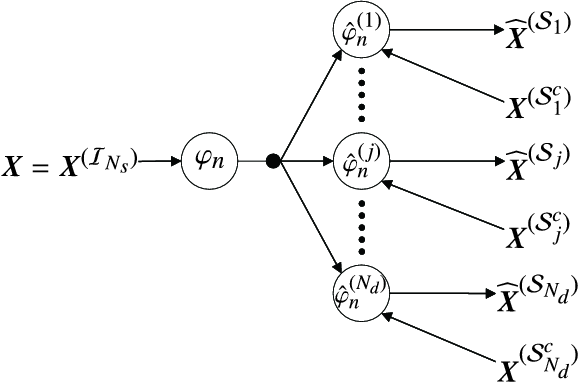}
    \caption{Generalized complementary delivery network}
    \label{fig:formulate}
  \end{center}
\end{figure}

Our main contributions in this paper include showing explicit constructions of universal
codes for other multiterminal source coding networks. Figs. \ref{fig:satellite:1} and
\ref{fig:satellite:2} illustrate the scenario we are considering: Several stations are
separately deployed in a field. Every station collects its own target data from sensors
or terminals, and wants to share all the target data with the other stations. To
accomplish
this task, each station transmits the collected data to a satellite, and the satellite
broadcasts all the received data back to the stations. Each station utilizes its own
target data as side information to reproduce all the other data. Willems et al.
\cite{BroadcastSatelliteCodingOrig,BroadcastSatelliteCoding} investigated a special case
of the above scenario in which three stations were deployed and each station had access
to one of three target messages, and they determined the minimum achievable rates for
uplink (from each station to the satellite) and downlink (from the satellite to all the
stations) transmissions. Their main result implies that the uplink transmission is
equivalent to the traditional Slepian-Wolf coding system \cite{SlepianWolf}, and thus
we should concentrate on the downlink part. Henceforth we denote the networks
characterized by the downlink transmission shown in Fig. \ref{fig:satellite:2} as
{\it generalized complementary delivery networks} (Fig. \ref{fig:formulate}), and
we denote the generalized complementary delivery network with two stations and
two target messages as the {\it (original) complementary delivery network}. This notation
is based on the network structure where each station (decoder) {\it complements} the
target messages from the codeword {\it delivered} by the satellite (encoder).

\begin{figure}[t]
  \begin{center}
    \includegraphics[width=0.5\hsize]{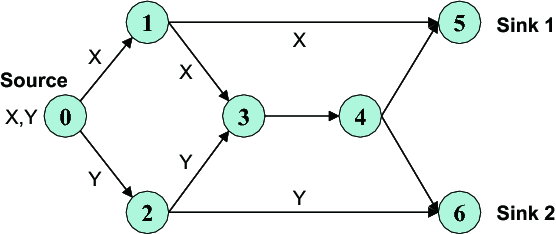}
    \caption{Butterfly network}
    \label{fig:butterfly}
  \end{center}
\end{figure}

The complementary delivery network can be regarded as a special example of the butterfly
network \cite{NetworkCodingOrig,LinearNetworkCoding} (Fig.\ref{fig:butterfly}), which is
one of a very well known network structure that represents the benefits of network
coding. If we assume that all the edges in Fig. \ref{fig:butterfly} except that between
nodes 3 and 4 have sufficiently large capacities, the problem is to find the minimum
capacities of the edge between the nodes 3 and 4 satisfying that allows two messages
emitted from the source (node 0) to be delivered to sinks 1 (node 5) and 2 (node 6). This
situation is equivalent to the complementary delivery network in which the messages
emitted from the source node are correlated with each other. Several coding problems for
correlated sources over a network have recently been investigated. At first only one
receiver was considered (e.g. \cite{NetworkSlepianWolfHan,NetworkCodingCorrelatedBarros})
, and later networks incorporating multiple receivers were studied (e.g.
\cite{SeparationCorrelatedNetwork,RandomLinearNetworkCoding,NetworkSlepianWolfCristescu,%
LinearUniversalComplementarySTW}). In particular, Ho et al.
\cite{RandomLinearNetworkCoding} and Kuzuoka et al.
\cite{LinearUniversalComplementarySTW} applied the linear Slepian-Wolf codes to random
linear network coding over general 2-source multi-cast networks and universal source
coding for the complementary delivery network, respectively. However, explicit code
constructions over networks with multiple sources and multiple destinations still remain
open.

This paper proposes a universal coding scheme for generalized complementary delivery
networks that involve multiple sources and multiple destinations. First, an explicit
construction of fixed-to-fixed length (FF) universal codes based on a graph-theoretical
analysis is presented. This construction utilizes a codebook expressed as a certain kind
of undirected graphs. Encoding can be regarded as the vertex coloring of the graphs. The
bounds of error probabilities and probabilities of correct decoding can be evaluated by
methods of types. The proposed coding scheme can always attain the optimal
error exponent (the exponent of error probabilities), and can attain the optimal exponent
of correct decoding in some cases. This FF coding scheme can be applied to
fixed-to-variable length (FV) universal codes. Overflow and underflow probabilities are
evaluated in almost the same way as the error probabilities and the
probabilities of correct decoding, respectively.

This paper is organized as follows: Notations and definitions are provided in Section
\ref{sec:pre}. A generic formulation of the generalized complementary delivery coding
system is introduced in Section \ref{sec:formulate}. A coding scheme for FF universal
codes is proposed in Section \ref{sec:construct}. Several coding theorems for FF
universal codes are clarified in Section \ref{sec:theorem}. Lastly, FV universal coding
is discussed in Section \ref{sec:variable}.

%%%%%%%%
\section{Preliminaries}
\label{sec:pre}

%%%%
\subsection{Basic definitions}
\label{sec:pre:defs}

%[Alphabets]
Let $\cB$ be a binary set, $\cB^*$ be the set of all finite sequences in the set $\cB$
and $\cI_M=\{1,2,\cdots,M\}$ for an integer $M$. In what follows, random variables are
denoted by capital letters such as $X$, and their sample values (resp. alphabets) by the
corresponding small letters (resp. calligraphic letters) such as $x$ (resp. $\cX$),
except as otherwise noted. The cardinality of a finite set $\cX$ is written as $|\cX|$,
and the $n$-th Cartesian product of $\cX$ by $\cX^n$.
%[Sequences]
A member of $\cX^n$ is written as 
\[
  x^n = (x_1,x_2,\cdots,x_n),
\]
and substrings of $x^n$ are written as
\[
  x_i^j = (x_i,x_{i+1},\cdots,x_j)\quad i\le j.
\]
When the dimension is clear from the context, vectors will be denoted by boldface
letters, i.e., $\x\in\cX^n$.

%[Probability distribution]
The probability distribution for a random variable $X$ is denoted by $P_X$. Similarly,
the probability distribution for random variables $(X,Y)$ is denoted by $P_{XY}$, and the
conditional distribution of $X$ given $Y$ is written as $P_{X|Y}$. The set of all
probability distributions on $\cX$ is written as $\cP(\cX)$, and the set of all
conditional distributions on $\cX$ given a distribution $P_Y\in\cP(\cY)$ is written as
$\cP(\cX|P_Y)$, which means that each member $P_{X|Y}$ of $\cP(\cX|P_Y)$ is characterized
by $P_{XY}\in\cP(\cX\times\cY)$ as $P_{XY}=P_{X|Y}P_Y$.
%[Sources]
A discrete memoryless source (DMS) is an infinite sequence of independent copies of a
random variable $X$. The alphabet of a DMS is assumed to be a finite set except as
otherwise noted. For simplicity, we denote a source $(\cX,P_{X})$ by referring to
its generic distribution $P_X$ or random variable $X$.
%[Multiple sources]
A set
\[ \X = (X^{(1)},X^{(2)},\cdots,X^{(N_s)}) \]
of $N_s$ random variables is also called a DMS, where each random variable $X^{(i)}$
takes a value in a finite set $\cX^{(i)}$ $(i\in\cI_{N_s})$. For a set
$\cS\subseteq\cI_{N_s}$, the corresponding subset of sources is written as
\begin{eqnarray*}
  \X^{(\cS)}  &\eqdef& \{X^{(i)}| i\in\cS\},
\end{eqnarray*}
and the corresponding subset of its sample sequences (resp. alphabets) $\cS$ is denoted
by
\begin{eqnarray*}
  \cX^{(\cS)} &\eqdef& \prod_{i\in\cS}\cX^{(i)},\\
  \x^{(\cS)}  &\eqdef& \{\x^{(i)}\in\cX^{(i)}| i\in\cS\}.
\end{eqnarray*}
For a set $\cS\subseteq\cI_{N_s}$, the $n$-th Cartesian product of $\cX^{(\cS)}$, its
member and the corresponding random variable are written as $\cX^{(\cS)n}$, $\x^{(\cS)n}$
and $\X^{(\cS)n}$, respectively. With $\cS=\cI_{N_s}$, we denote
$\X^{(\cS)n}=\X^n$. For a set $\cS\subseteq\cI_{N_s}$, its complement is denoted as
$\cS^c=\cI_{N_s}-\cS$.

%[Entropy, mutual information, divergence]
For a DMS $\X$ and finite sets $\cS_1, \cS_2\subseteq\cI_{N_s}$ that satisfy
$\cS_1\cap\cS_2=\emptyset$, the joint entropy of $\X^{(\cS_1)}$ and the conditional
entropy of $\X^{(\cS_2)}$ given $\X^{(\cS_1)}$ are written as $H(\X^{(\cS_1)})$ and
$H(\X^{(\cS_2)}|\X^{(\cS_1)})$, respectively (cf. \cite{CsiszarKorner}). For a generic
distribution $P\in\cP(\cX^{(\cS_1)})$ and a conditional distribution
$W\in\cP(\cX^{(\cS_2)}|P)$, $H(P)$ and $H(W|P)$ also represent the joint entropy of
$\X^{(\cS_1)}$ and the conditional entropy of $\X^{(\cS_2)}$ given $\X^{(\cS_1)}$, where
$P=P_{\X^{(\cS_1)}}$ and $W=P_{\X^{(\cS_2)}|\X^{(\cS_1)}}$. The Kullback-Leibler
divergence, or simply the divergence, between two distributions $P$ and $Q$ is written as
$D(P\|Q)$.

%[Misc]
In the following, all bases of exponentials and logarithms are set at 2.

%%%%
\subsection{Types of sequences}
\label{sec:pre:type}

Let us define the {\it type} of a sequence $\x\in\cX^n$ as the empirical distribution
$Q_{\x}\in\cP(\cX)$ of the sequence $\x$, i.e.
\begin{eqnarray*}
  Q_{\x}(a) &\eqdef& \frac 1n N(a|\x)\quad\forall a\in\cX,
\end{eqnarray*}
where $N(a|\x)$ represents the number of occurrences of the letter $a$ in the sequence
$\x$. Similarly, the joint type $Q_{\x^{(\cS)}}\in\cP(\cX^{\cS})$ for a given set
$\cS\subseteq\cI_{N_s}$ is defined by
\begin{eqnarray*}
  \lefteqn{Q_{\x^{(\cS)}}(a_{i_1},a_{i_2},\cdots,a_{i_{|\cS|}})}\\
  &\eqdef& \frac 1n N(a_{i_1},a_{i_2},\cdots,a_{i_{|\cS|}}|\x^{(\cS)})\\
  & & \forall(a_{i_1},a_{i_2},\cdots,a_{i_{|\cS|}})\in\cX^{(\cS)}.
\end{eqnarray*}
Let $\cP_n(\cX)$ be the set of types of sequences in $\cX^n$. Similarly, for every type
$Q\in\cP_n(\cX)$, let $\cV_n(\cY|Q)$ be the set of all stochastic matrices $V:\cX\to\cY$
such that for some pairs $(\x,\y)\in\cX^n\times\cY^n$ of sequences we have
\[
  Q_{\x,\y}(\x,\y) = Q(\x)V(\y|\x) = \prod_{i=1}^n Q(x_i)V(y_i|x_i).
\]
For every type $Q\in\cP_n(\cX)$ we denote
\[
  T_Q^n ~{\displaystyle\mathop{=}^{\mbox{\rm def.}}}~
    \{\x\in\cX^n| Q_{\x}=Q\}.
\]
Similarly, for every sequence $\x\in T_Q^n$ and stochastic matrix $V\in\cV_n(\cY|Q)$, we
define a {\it V-shell} as
\begin{eqnarray*}
  \lefteqn{T_V^n(\x) {\displaystyle\mathop{=}^{\mbox{\rm def.}}}}\\
  && \{\y\in\cY^n| Q(x)V(y|x)=Q_{\x,\y}(x,y),~\forall(x,y)\in\cX\times\cY\}.
\end{eqnarray*}

Here, let us introduce several important properties of types.

\begin{hodai} \label{lemma:typecount}
  {\rm (Type counting lemma \cite[Lemma~2.2]{CsiszarKorner})}
  \[ |\cP_n(\cX)|\le (n+1)^{|\cX|}. \]
\end{hodai}

\begin{hodai} \label{lemma:sizeshell}
  {\rm (Sizes of V-shells \cite[Lemma~2.5]{CsiszarKorner})}\\
  For every type $Q\in\cP_n(\cX)$, sequence $\x\in T_Q^n$ and stochastic matrix
  $V: \cX\to\cY$ such that $T_V^n(\x)\neq\emptyset$, we have
  \begin{eqnarray*}
    |T_V^n(\x)| &\ge& (n+1)^{-|\cX||\cY|}\exp\{nH(V|Q)\},\\
    |T_V^n(\x)| &\le& \exp\{nH(V|Q)\}.
  \end{eqnarray*}
\end{hodai}

\begin{hodai} \label{lemma:prob}
  {\rm (Probabilities of types \cite[Lemma~2.6]{CsiszarKorner})}\\
  For every type $Q\in\cP_n(\cX)$ and every distribution $P_X\in\cP(\cX)$, we have
  \begin{eqnarray*}
    P_X(\x)  & = & \exp\{-n(D(Q\| P_X)+H(Q))\}\quad\forall\x\in T_Q,\\
    P_X(T_Q) &\ge& (n+1)^{-|\cX|}\exp\{-nD(Q\| P_X)\},\\
    P_X(T_Q) &\le& \exp\{-nD(Q\| P_X)\}.
  \end{eqnarray*}
\end{hodai}

%%%%
\subsection{Graph coloring}
\label{sec:pre:graph}

Let us introduce several notations and lemmas related to graph coloring.
%[Basic notations]
A (undirected) graph is denoted as $G=(V_G,E_G)$, where $V_G$ is a set of vertices and
$E_G$ is a set of edges. The degree $\Delta(v)$ of a vertex $v\in V_G$ is the number of
other vertices connected by edges, and the degree $\Delta(G)$ of a graph is the
maximum number of degrees of vertices in the graph $G$. A graph where an edge connects
every pair of vertices is called a {\it complete graph}. A complete subgraph is called
a {\it clique}, and the largest degree of cliques in a graph $G$ is called the
{\it clique number} $\omega(G)$ of the graph $G$.
%[coloring]
The {\it vertex coloring}, or simply {\it coloring} of a graph $G$ is where no two
adjacent vertices are assigned the same symbol. The number of symbols necessary for the
vertex coloring of a graph is called the {\it chromatic number} $\chi(G)$. Similarly, the
{\it edge coloring} of a graph $G$ is where no two adjacent edges are assigned the same
symbol, and the number of symbols necessary for edge coloring is called the
{\it edge chromatic number} $\chi'(G)$.

%[Coloring lemmas]
The following lemmas are well known as bounds of the chromatic number and the edge
chromatic number.

\begin{hodai} \label{lemma:brooks}
  {\rm (Brooks \cite{chromaticnumber:brooks,GraphTheory:biggs})}
  \begin{eqnarray*}
    && \omega(G) \le \chi(G) \le \Delta(G)
  \end{eqnarray*}
  unless $G$ is a complete graph or an odd cycle (a cycle graph that contains an odd
  number of vertices).
\end{hodai}

\begin{hodai} \label{lemma:vizing}
  {\rm (Vizing \cite{chromaticnumber:vizing,GraphTheory:biggs})}
  \begin{eqnarray*}
    && \Delta(G) \le \chi'(G) \le \Delta(G)+1.
  \end{eqnarray*}
\end{hodai}

\begin{hodai} \label{lemma:graph:konig}
  {\rm (K\dotdot{o}nig \cite{EdgeColoring,GraphTheory:biggs})}\\
  If a graph $G$ is bipartite, then
  \begin{eqnarray*}
    && \chi'(G)=\Delta(G).
  \end{eqnarray*}
\end{hodai}

%%%%%%%%
\section{Problem formulation}
\label{sec:formulate}

This section formulates the coding problem investigated in this paper, and shows the
fundamental bound of the coding rate.

First, we describe a generalized complementary delivery network. Fig. \ref{fig:formulate}
represents the network formulated below. This network is composed of $N_s$ sources
$\X=\X^{(\cI_{N_s})}$, one encoder $\varphi_n$ and $N_d$ decoders $\wh{\varphi}_n^{(1)}$
$\cdots\wh{\varphi}_n^{(N_d)}$. Each decoder $\wh{\varphi}_n^{(j)}$ has access to side
information $\X^{(\cS_j^c)}$ $(\cS_j\subset\cI_{N_s})$ to enable it to reproduce the
information $\X^{(\cS_j)}$. Since the indices $\bcS=\{\cS_j\}_{j=1}^{N_d}$ of side
information determine the network, henceforth we denote the network by $\bcS$. Without
loss of generality, we assume $\cS_{j_1}\neq\cS_{j_2}$ $\forall j_1,j_2\in\cI_{N_d}$.

Based on the above definition of the network, we formulate the coding problem for the
network.

\begin{teigi}  \label{def:code}
  {\rm (Fixed-to-fixed generalized complementary delivery (FF-GCD) code)}\\
  A sequence
  \[
    \{( \varphi_n,\wh{\varphi}_n^{(1)},\cdots,\wh{\varphi}_n^{(N_d)} )\}_{n=1}^{\infty}
  \]
  of codes
  \[
    ( \varphi_n,\wh{\varphi}_n^{(1)},\cdots,\wh{\varphi}_n^{(N_d)} )
  \]
  is an FF-GCD code for the network $\bcS=\{\cS_j\}_{j=1}^{N_d}$ if
  \begin{eqnarray*}
    \varphi_n            &:& \cX^{(\cI_{N_s})n}\rightarrow\cI_{M_n}\\
    \wh{\varphi}_n^{(j)} &:& \cI_{M_n}\times\cX^{(\cS_j^c)n}\rightarrow\cX^{(\cS_j)n}
                             \quad\forall j\in\cI_{N_d}.
  \end{eqnarray*}
\end{teigi}

\begin{teigi}  \label{def:rate}
  {\rm(FF-GCD achievable rate)}\\
  $R$ is an FF-GCD achievable rate of the source $\X$ for the network $\bcS$ if and only
  if there exists an FF-GCD code
  \[
    \{( \varphi_n,\wh{\varphi}_n^{(1)},\cdots,\wh{\varphi}_n^{(N_d)} )\}_{n=1}^{\infty}
  \]
  for the network $\bcS$ that satisfies
  \begin{eqnarray*}
    \limsup_{n\to\infty}\frac 1n\log M_n &\le& R,\\
    \lim_{n\to\infty}e_n^{(j)}           & = & 0\quad\forall j\in\cI_{N_d}.
  \end{eqnarray*}
  where
  \begin{eqnarray*}
    e_n^{(j)} &=& \Pr\left\{\X^{(\cS_j)n}\neq\wh{\X}^{(\cS_j)n}\right\}
                  \quad\forall j\in\cI_{N_d},\\
    \wh{\X}^{(\cS_j)n} &{\displaystyle\mathop{=}^{\mbox{\rm def.}}}&
      \wh{\varphi}_n^{(j)}(\varphi_n(\X^n),\X^{(\cS_j^c)n}).
  \end{eqnarray*}
\end{teigi}

\begin{teigi}  \label{def:min_rate}
  {\rm(Inf FF-GCD achievable rate)}
  \begin{eqnarray*}
    \lefteqn{R_f(\X|\bcS)}\\
    &=& \inf\{R|
        R\mbox{ is an FF-GCD achievable rate of }\X\mbox{ for }\bcS\}.
  \end{eqnarray*}
\end{teigi}

\begin{figure}[t]
  \begin{center}
    \includegraphics[width=0.5\hsize]{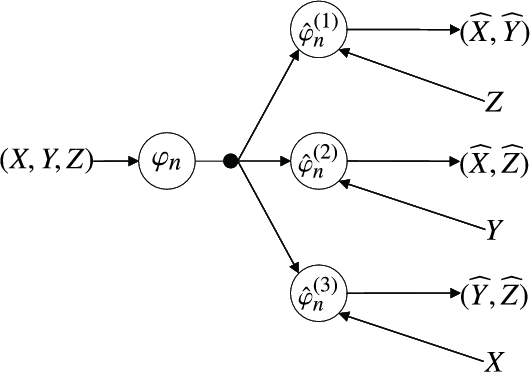}
    \caption{Network investigated by Willems et al.}
    \label{fig:complementary:three}
  \end{center}
\end{figure}

Willems et al. \cite{BroadcastSatelliteCodingOrig,BroadcastSatelliteCoding} clarified the
minimum achievable rate $R_f(\X|\bcS)$ for a special case, where $N_s=N_d=3$,
$(X_1,X_2,X_3)=(X,Y,Z)$, $\cS_1=\{1,2\}$, $\cS_2=\{1,3\}$ and $\cS_3=\{2,3\}$ (Fig.
\ref{fig:complementary:three}).
\begin{teiri}  \label{theorem:willems}
  {\rm (Coding theorem of FF-GCD codes for three users
    \cite{BroadcastSatelliteCoding})}\\
  If $N_s=N_d=3$, $(X_1,X_2,X_3)=(X,Y,Z)$, $\cS_1=\{1,2\}$, $\cS_2=\{1,3\}$ and
  $\cS_3=\{2,3\}$, then
  \begin{eqnarray*}
    \lefteqn{R_f(X,Y,Z|\bcS)}\\
    &=& \max\{H(X,Y|Z),H(Y,Z|X),H(X,Z|Y)\}
  \end{eqnarray*}
\end{teiri}

It is easy to extend Theorem \ref{theorem:willems} to the following coding theorem for
general cases:
\begin{teiri}  \label{theorem:general}
  {\rm (Coding theorem of FF-GCD codes for general cases)}
  \begin{eqnarray*}
    R_f(\X|\bcS)
    &=& \max_{j\in\cI_{N_d}}
        H\left(\X^{(\cS_j)}\left|\X^{(\cS_j^c)}\right.\right)
  \end{eqnarray*}
\end{teiri}

\begin{hosoku}
  The generalized complementary delivery network is included in the framework considered
  by Csisz\dash{a}r and K\dotdot{o}rner \cite{UniversalSlepianWolf:Csiszar}. Therefore,
  Theorem \ref{theorem:general} can be obtained as a corollary of their results.
\end{hosoku}

%%%%%%%%
\section{Code construction}
\label{sec:construct}

This section shows an explicit construction of universal codes for the generalized
complementary delivery network. The proposed universal coding scheme is described as
follows:

\begin{figure}[t]
  \begin{center}
    \includegraphics[width=0.25\hsize]{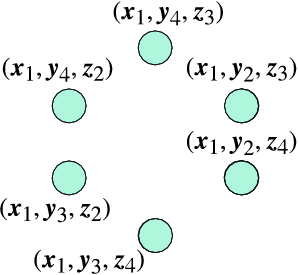}\hspace{3mm}
    \includegraphics[width=0.25\hsize]{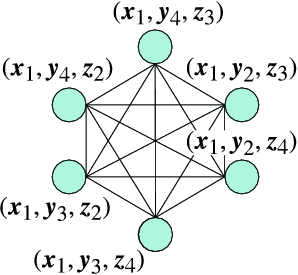}\\\vspace{7mm}
    \includegraphics[width=0.25\hsize]{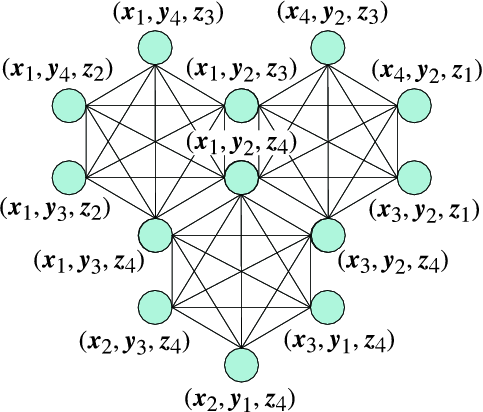}\hspace{3mm}
    \includegraphics[width=0.25\hsize]{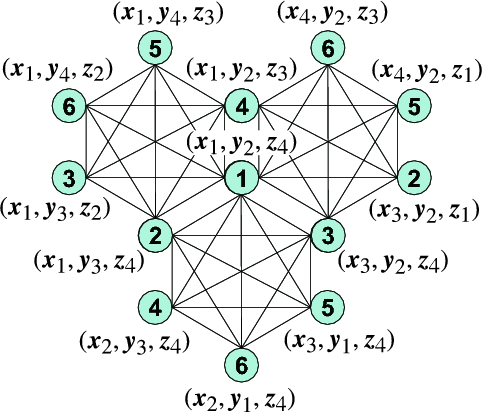}
    \caption{
      (Upper left) Intuitive example of coding graph. Each node corresponds to a sequence
      set $(\x_i,\y_j,\z_k)\in T_{Q_{XYZ}}^n$.
    }
    \label{fig:universal:CodingGraphNew1}
    \caption{
      (Upper right) For a given $\x_1$, an edge is placed between every pair of vertices
      whose subsequences satisfy $(\y_j,\z_k)\in T_{V_3}^n(\x_1)$, which means that for a
      given $\x_1$ we must distinguish each $(\y_j,\z_k)$ such that
      $(\x_1,\y_j,\z_k)\in T_{Q_{XYZ}}^n$.
    }
    \label{fig:universal:CodingGraphNew2}
    \caption{
      (Lower left) In a similar manner, for a given $\y_2$ (resp. $\z_4$) an edge is
      deployed between every pair of vertices whose subsequences satisfy
      $(\x_i,\z_k)\in T_{V_2}^n(\y_2)$ (resp. $(\x_i,\y_j)\in T_{V_1}^n(\z_4)$).
    }
    \label{fig:universal:CodingGraphNew3}
    \caption{
      (Lower right) Example of codeword assignment. Assigning a codeword to each sequence
      set can be regarded as vertex coloring of the coding graph.
    }
    \label{fig:universal:CodingGraphNew4}
  \end{center}
\end{figure}

\medskip\noindent
[Encoding]
\begin{enumerate}
  \item
    Determine a set $\cT_n(R)\subseteq\cP_n(\cX^{(\cI_{N_s})})$ of joint types as
    \begin{eqnarray*}
      \lefteqn{\cT_n(R) = \{Q_{\X}\in\cP_n(\cX^{(\cI_{N_s})}):}\\
      & & \hspace{-5mm}\max_{j\in\cI_{N_d}}\{H(V_j|Q_j)\}\le R,
      \quad Q_{\X}=Q_jV_j,\\
      & & \hspace{-5mm}Q_j\in\cP_n(\cX^{(\cS_j^c)}),
          V_j\in\cV_n(\cX^{(\cS_j)}|Q_j), \forall j\in\cI_{N_d}\},
    \end{eqnarray*}
    where $R>0$ is a given coding rate. We note that the joint type $Q_{\X}$ and the
    system $\bcS$ specify the type $Q_j$ and the conditional type $V_j$ for every
    $j\in\cI_{N_d}$.
  \item
    Create a graph for every joint type $Q_{\X}\in\cT_n(R)$. An intuitive example of
    coding graphs is shown in Figs. \ref{fig:universal:CodingGraphNew1},
    \ref{fig:universal:CodingGraphNew2}, \ref{fig:universal:CodingGraphNew3} and
    \ref{fig:universal:CodingGraphNew4}, where the network shown in Fig.
    \ref{fig:complementary:three} is considered. Each vertex of the graph corresponds to
    a sequence set $\x^{(\cI_{N_s})}\in T_{Q_{\X}}^n$ (cf. Fig.
    \ref{fig:universal:CodingGraphNew1}). Henceforth we denote a vertex by referring to
    the corresponding sequence set $\x^{(\cI_{N_s})}$. An edge is placed between vertices
    $\x_1^{(\cI_{N_s})}$ and $\x_2^{(\cI_{N_s})}$ if and only if
    $\x_1^{(\cS_j^c)}=\x_2^{(\cS_j^c)}$, $\x_1^{(\cS_j)}\in T_{V_j}^n(\x_1^{(\cS_j^c)})$
    and $\x_2^{(\cS_j)}\in T_{V_j}^n(\x_2^{(\cS_j^c)})$ for some $j\in\cI_{N_d}$ (cf.
    Figs. \ref{fig:universal:CodingGraphNew2} and \ref{fig:universal:CodingGraphNew3}).
    In the following, we call this graph the {\it coding graph} $G(Q_{\X})$. Note that
    Figs. \ref{fig:universal:CodingGraphNew3} and \ref{fig:universal:CodingGraphNew4}
    show only a subgraph that corresponds to V-shells $T_{V_j}^n(\x^{(\cS_j^c)})$, where
    $\x^{(\cS_1^c)}=\x_1$, $\x^{(\cS_2^c)}=\y_2$ and $\x^{(\cS_3^c)}=\z_4$.
  \item
    Assign a symbol to each vertex of the coding graph $G(Q_{\X})$ so that the same
    symbol is not assigned to any pairs of adjacent vertices (cf. Fig.
    \ref{fig:universal:CodingGraphNew4}).
  \item
    For an input sequence set $\x^{(\cI_{N_s})}$ whose joint type $Q_{\X}$ is a member of
    $\cT_n(R)$, the index assigned to the joint type $Q_{\X}$ is the first part of the
    codeword, and the symbol assigned to the corresponding vertex of the coding graph is
    determined as the second part of the codeword. For a sequence set $\x^{(\cI_{N_s})}$
    whose joint type $Q_{\X}$ is not a member of $\cT_n(R)$, the codeword is determined
    arbitrarily and an encoding error is declared.
\end{enumerate}

\noindent
[Decoding: $\wh{\varphi}_n^{(j)}$]
\begin{enumerate}
  \item
    The first part of the received codeword represents the joint type $\wh{Q}_{\X}$ of
    the input sequence. If no encoding error occurs, then $\wh{Q}_{\X}$ should be
    $Q_{\X}$, and therefore the decoder $\wh{\varphi}_n^{(j)}$ can find the coding graph
    $\wh{G}(\wh{Q}_{\X})=G(Q_{\X})$ used in the encoding scheme.
  \item
    For given side information $\x_1^{(\cS_j^c)}$ and the joint type $Q_{\X}$,
    find the vertex $\x_2^{(\cI_{N_s})}$ such that (i)
    $\x_2^{(\cS_j^c)}=\x_1^{(\cS_j^c)}$ and (ii) the second part of the received codeword
    is assigned to $\x_2^{(\cI_{N_s})}$. Such a vertex is found in the clique that
    corresponds to the set $T_{V_j}^n(\x^{(\cS_j^c)})$. With Fig.
    \ref{fig:universal:CodingGraphNew4}, if $\x^{(\cS_1^c)}=\x_1$ is given as a side
    information sequence, we can find such a vertex from the upper left clique. Note that
    the conditional type $V_j$ has been determined by $\wh{Q}_{\X}=Q_{\X}$. The sequence
    set $\wh{\x}^{(\cS_j)}\in T_{Q_j}^n$ found in this step is reproduced.
\end{enumerate}

It should be noted that the above coding scheme is universal since it does not depend
on the distribution $P_{\X}$ of a source $\X$.

The coding rate of the above proposed coding scheme is determined by the chromatic number
of the coding graph $G(Q_{\X})$. To this end, we introduce the following lemmas.

\begin{hodai} \label{lemma:graph:property}
  The coding graph $G(Q)$ of the joint type $Q=Q_{\X}$ has the following
  properties:
  \begin{enumerate}
    \item
      Every vertex set
      \[ T_{V_j}^n(\x^{(\cS_j^c)}) \quad(j\in\cI_{N_d})\]
      comprises a clique, where
      \begin{eqnarray*}
	&& Q=Q_jV_j, \quad\x^{(\cS_j^c)}\in T_{Q_j}^n,\\
	&& Q_j\in\cP_n(\cX^{(\cS_j^c)}) \quad V_j\in\cV_n(\cX^{(\cS_j)}|Q_j).
      \end{eqnarray*}
    \item
      Every vertex $\x^{(\cI_{N_s})}\in T_Q^n$ belongs to $N_d$ cliques, each of which
      corresponds to the vertex set
      \[ T_{V_j}^n(\x^{(\cS_j^c)}). \quad(j\in\cI_{N_d})\]
    \item
      The vertex $\x^{(\cI_{N_s})}\in T_Q^n$ has no edges from vertices not included in
      the vertex sets $\cup_{j\in\cI_{N_d}}T_{V_j}^n(\x^{(\cS_j^c)})$.
    \item
      For a given joint type $Q\in\cP_n(\cX^{(\cI_{N_s})})$, both the clique number
      $\omega(G(Q))$ and the degree $\Delta(G(Q))$ of the coding graph $G(Q)$ are
      constant and obtained as follows:
      \begin{eqnarray*}
	\omega(G(Q)) & = & \max_{j\in\cI_{N_d}}|T_{V_j}^n(\x^{(\cS_j^c)})|,\\
	\Delta(G(Q)) & = & \sum_{j\in\cI_{N_d}}|T_{V_j}^n(\x^{(\cS_j^c)})|.
      \end{eqnarray*}
  \end{enumerate}
\end{hodai}

\begin{IEEEproof}
1) 2) 3) Easily obtained from the first and second steps of the above encoding scheme.
4) Easily obtained from the above properties.
\end{IEEEproof}

\begin{hodai} \label{lemma:graph:chromatic}
  The chromatic number of the coding graph $G(Q)$ of the joint type
  $Q\in\cT_n(R)$ is bounded as
  \begin{eqnarray*}
    \chi(G(Q)) &\le& N_d\exp(nR).
  \end{eqnarray*}
\end{hodai}

\begin{IEEEproof}
  This property is directly derived from Lemmas \ref{lemma:sizeshell}, \ref{lemma:brooks}
  and \ref{lemma:graph:property} as follows:
  \begin{eqnarray}
    \chi(G(Q))
    &\le& \Delta(G(Q)) \label{eq:proof:chromatic:1}\\
    & = & \sum_{j\in\cI_{N_d}} |T_{V_j}^n(\x^{(\cS_j^c)})|
          \label{eq:proof:chromatic:2}\\
    &\le& \sum_{j\in\cI_{N_d}}\exp\{nH(V_j|Q_j)\} \label{eq:proof:chromatic:3}\\
    &\le& N_d\exp\{n\max_{j\in\cI_{N_d}}H(V_j|Q_j)\} \nonumber\\
    &\le& N_d\exp(nR). \label{eq:proof:chromatic:4}
  \end{eqnarray}
  where Eq. (\ref{eq:proof:chromatic:1}) comes from Lemma \ref{lemma:brooks}, 
  Eq. (\ref{eq:proof:chromatic:2}) from Lemma \ref{lemma:graph:property},
  Eq. (\ref{eq:proof:chromatic:3}) from Lemma \ref{lemma:sizeshell}, and
  Eq. (\ref{eq:proof:chromatic:4}) from the definition of $\cT_n(R)$.
  This concludes the proof of Lemma \ref{lemma:graph:chromatic}.
\end{IEEEproof}

\medskip
From the above discussions, we obtain
\[
  \omega(G(Q)) \le \chi(G(Q)) \le \Delta(G(Q)) \le N_d\exp(nR).
\]

%%%%%%%%
\section{Coding theorems}
\label{sec:theorem}

%%%%%%%%
\subsection{General cases}
\label{sec:theorem:general}

We show several coding theorems derived from the proposed coding scheme. Before showing
these coding theorems, let us define the following function:
\begin{eqnarray}
  \epsilon_n(N)
  &{\displaystyle\mathop{=}^{\mbox{\rm def.}}}&
  \frac 1n \{|\cX^{(\cI_{N_s})}|\log(n+1)+\log N\}\label{eq:theorem:direct:2}\\
  &\to& 0\quad(n\to\infty). \nonumber
\end{eqnarray}

First we present the direct part of the coding theorem for the universal FF-GCD codes,
which implies that the coding scheme shown in Section \ref{sec:construct} attains the
minimum achievable rate.

\begin{teiri} \label{theorem:UniversalCode:direct}
  For a given real number $R>0$, there exists a universal FF-GCD code
  \[
    \{( \varphi_n,\wh{\varphi}_n^{(1)},\cdots,\wh{\varphi}_n^{(N_d)} )\}_{n=1}^{\infty}
  \]
  for the network $\bcS$ such that for any integer $n\ge 1$ and any source $\X$
  \begin{eqnarray}
    \frac 1n\log M_n &\le& R+\epsilon_n(N_d), \label{eq:theorem:direct:1}\\
    \sum_{j=1}^{N_d}e_n^{(j)} &\le& \nonumber\\
    & & \hspace{-16mm}
    \exp\left\{-n\left(-\epsilon_n(N_d)+\hspace{-2mm}\min_{Q_{\X}\in\cT_n^c(R)}
    \hspace{-2mm}D(Q_{\X}\| P_{\X})\right)\right\}. \nonumber
  \end{eqnarray}
\end{teiri}

\begin{IEEEproof}
Note that a codeword is composed of two parts: the first part corresponds to the joint
type of an input sequence set, and the second part represents a symbol assigned to the
input sequence set in the coding graph of the joint type. Therefore, the size of the
codeword set is bounded as
\begin{eqnarray*}
  M_n &\le& |\cP_n(\cX^{(\cI_{N_s})})|\cdot N_d\exp(nR)\\
      &\le& N_d(n+1)^{|\cX^{(\cI_{N_s})}|}\exp(nR),
            \quad(\mbox{Lemma \ref{lemma:typecount}})
\end{eqnarray*}
which implies Eq. (\ref{eq:theorem:direct:1}). Next, we evaluate decoding error
probabilities. Since every sequence set $\x^{(\cI_{N_s})n}$ whose joint type is a member
of $\cT_n(R)$ is reproduced correctly at the decoder, the sum of the error probabilities
is bounded as
\begin{eqnarray}
  \lefteqn{\sum_{j=1}^{N_d}e_n^{(j)}}\\
  &\le& N_d\Pr\left\{\X^n\in T_{\wt{Q}_{\X}}^n: \wt{Q}_{\X}\in\cT_n^c(R)\right\}
        \nonumber\\
  &\le& N_d\sum_{\wt{Q}_{\X}\in\cT_n^c(R)}\exp\{-nD(\wt{Q}_{\X}\| P_{\X})\}
	\label{eq:proof:direct:1}\\
  &\le& N_d\sum_{\wt{Q}_{\X}\in\cT_n^c(R)}\hspace{-4mm}
        \exp\left\{-n\min_{Q_{\X}\in\cT_n^c(R)}D(Q_{\X}\| P_{\X})\right\}\nonumber\\
  &\le& N_d(n+1)^{|\cX^{(\cI_{N_s})}|} \nonumber\\
  &   & \hspace{10mm}\times\exp\left\{-n\min_{Q_{\X}\in\cT_n^c(R)}
	D(Q_{\X}\| P_{\X})\right\} \label{eq:proof:direct:2}\\
  & = & \exp\left\{-n\left(-\epsilon_n(N_d)+\hspace{-2mm}\min_{Q_{\X}\in\cT_n^c(R)}
	D(Q_{\X}\| P_{\X})\right)\right\},\nonumber
\end{eqnarray}
where Eq. (\ref{eq:proof:direct:1}) comes from Lemma \ref{lemma:prob}, and
Eq. (\ref{eq:proof:direct:2}) from Lemma \ref{lemma:typecount}.
This completes the proof of Theorem \ref{theorem:UniversalCode:direct}.
\end{IEEEproof}

\medskip
We can see that for any real value $R\ge R_f(\X|\bcS)$ we have
\begin{eqnarray*}
  \min_{Q_{\X}\in\cT_n^c(R)}D(Q_{\X}\|P_{\X}) &>& 0.
\end{eqnarray*}
This implies that if $R\ge R_f(\X|\bcS)$ there exists an FF-GCD code for the network
$\bcS$ that universally attains the conditions shown in Definition \ref{def:rate}.

The following converse theorem indicates that the error exponent obtained in Theorem
\ref{theorem:UniversalCode:direct} is tight.

\begin{teiri} \label{theorem:UniversalCode:converse}
  Any FF-GCD code
  \[
    \{( \varphi_n,\wh{\varphi}_n^{(1)},\cdots,\wh{\varphi}_n^{(N_d)} )\}_{n=1}^{\infty}
  \]
  for the system $\bcS$ must satisfy
  \begin{eqnarray*}
    \lefteqn{\sum_{j=1}^{N_d}e_n^{(j)}}\\
    &\ge& \exp\left\{-n\left(\epsilon_n(2)+\hspace{-2mm}
          \min_{Q_{\X}\in\cT_n^c(R+\epsilon_n(2))}D(Q_{\X}\| P_{\X})\right)\right\}
  \end{eqnarray*}
  for any integer $n\ge 1$, any source $\X$ and a given coding rate $R=1/n\log M_n>0$.
\end{teiri}

\begin{IEEEproof}
Note that the number of sequences to be decoded correctly for each decoder is at most
$\exp(nR)$. Here, let us consider a joint type $Q_{\X}\in\cT_n^c(R+\epsilon_n(2))$.
The definition of $\cT_n^c(R+\epsilon_n(2))$ and Lemma \ref{lemma:sizeshell} imply
that for $\x^{(\cI_{N_s})}\in T_{Q_{\X}}^n$ we have
\begin{eqnarray}
  \lefteqn{\max_{j\in\cI_{N_d}}\{|T_{V_j}^n(\x^{(\cS_j^c)})|\}}\nonumber\\
  &\ge& (n+1)^{-|\cX^{(\cI_{N_s})}|}\max_{j\in\cI_{N_d}}\exp\{nH(V_j|Q_j)\}
        \label{eq:proof:converse:1}\\
  &\ge& (n+1)^{-|\cX^{(\cI_{N_s})}|}\exp\{n(R+\epsilon_n(2))\}
        \label{eq:proof:converse:2}\\
  & = & 2\exp(nR),\nonumber
\end{eqnarray}
where Eq. (\ref{eq:proof:converse:1}) comes from Lemma \ref{lemma:sizeshell}, and Eq.
(\ref{eq:proof:converse:2}) from the definition of $\cT_n^c(R+\epsilon_n(2))$. Therefore,
at least half of the sequence sets in $T_{Q_{\X}}^n$ will not be decoded correctly at the
decoder $\wh{\varphi}_n^{(j)}$. Thus, the sum of the error probabilities is bounded as
\begin{eqnarray}
  \lefteqn{\sum_{j\in\cI_{N_d}}e_n^{(j)}}\nonumber\\
  &\ge& \frac{1}{2}\sum_{Q_{\X}\in\cT_n^c(R+\epsilon_n(2))}
        \Pr\{\X^n\in T_{Q_{\X}}^n\} \nonumber\\
  &\ge& \frac{1}{2}(n+1)^{-|\cX^{(\cI_{N_s})}|}
	\hspace{-5mm}\sum_{Q_{\X}\in\cT_n^c(R+\epsilon_n(2))}
	\hspace{-5mm}\exp\{-nD(Q_{\X}\| P_{\X})\} \nonumber\\
  &   & \label{eq:proof:converse:3}\\
  &\ge& \frac{1}{2}(n+1)^{-|\cX^{(\cI_{N_s})}|} \nonumber\\
  &   & \times\exp\left\{-n\hspace{-2mm}\min_{Q_{\X}\in\cT_n^c(R+\epsilon_n(2))}
	\hspace{-2mm}D(Q_{\X}\| P_{\X})\right\} \nonumber\\
  & = & \exp\left\{-n\left(\epsilon_n(2)+\hspace{-2mm}\min_{Q_{\X}\in
	\cT_n^c(R+\epsilon_n(2))}D(Q_{\X}\| P_{\X})\right)\right\}, \nonumber
\end{eqnarray}
where Eq. (\ref{eq:proof:converse:3}) comes from Lemma \ref{lemma:prob}.
This concludes the proof of Theorem \ref{theorem:UniversalCode:converse}.
\end{IEEEproof}

\medskip
The following corollary is directly derived from Theorems
\ref{theorem:UniversalCode:direct} and \ref{theorem:UniversalCode:converse}. This shows
the asymptotic optimality of the proposed coding scheme.

\begin{kei}
  For a given real number $R>0$, there exists a universal FF-GCD code
  \[
    \{( \varphi_n,\wh{\varphi}_n^{(1)},\cdots,\wh{\varphi}_n^{(N_d)} )\}_{n=1}^{\infty}
  \]
  for the network $\bcS$ such that for any source $\X$ 
  \begin{eqnarray*}
    && \limsup_{n\to\infty}\frac 1n\log M_n \le R,\\
    && \lim_{n\to\infty}-\frac 1n\log\sum_{j\in\cI_{N_d}}e_n^{(j)}
       = \min_{Q_{\X}\in\cT^c(R)}D(Q_{\X}\| P_{\X}),
  \end{eqnarray*}
  where
  \begin{eqnarray*}
    \lefteqn{\cT(R) = \{Q_{\X}\in\cP(\cX^{(\cI_{N_s})}):}\\
    & & \hspace{2mm}\max_{j\in\cI_{N_d}}H(V_j|Q_j)\le R,
        \quad Q_{\X}=Q_jV_j,\\
    & & \hspace{2mm}Q_j\in\cP_n(\cX^{(\cS_j^c)}),
        V_j\in\cV(\cX^{(\cS_j)}|Q_j), \forall j\in\cI_{N_d}\}.
  \end{eqnarray*}
\end{kei}

In a similar manner, we can obtain a probability such that the original sequence set
is correctly reproduced. The following theorem shows the lower bound of the probability
of correct decoding that can be achieved by the proposed coding scheme.

\begin{teiri} \label{theorem:UniversalCode:lowrate:direct}
  For a given real number $R>0$, there exists a universal FF-GCD code
  \[
    \{( \varphi_n,\wh{\varphi}_n^{(1)},\cdots,\wh{\varphi}_n^{(N_d)} )\}_{n=1}^{\infty}
  \]
  for the network $\bcS$ such that for any integer $n\ge 1$ and any source $\X$ 
  \begin{eqnarray}
    \frac 1n\log M_n &\le& R+\epsilon_n(N_d), \label{eq:theorem:lowrate:direct:1}\\
    1-\sum_{j=1}^{N_d}e_n^{(j)}
    &\ge& \nonumber\\
    &   & \hspace{-13mm}\exp\left\{-n\left(\epsilon_n(1)+\min_{Q_{\X}\in\cT_n(R)}
          D(Q_{\X}\| P_{\X})\right)\right\}.\nonumber
  \end{eqnarray}
\end{teiri}

\begin{IEEEproof}
Eq. (\ref{eq:theorem:lowrate:direct:1}) is derived in the same way as the proof of
Theorem \ref{theorem:UniversalCode:direct}. Next, we evaluate the probability such that
the original sequence set is correctly reproduced. Since every sequence set
$\x^{(\cI_{N_s})}$ whose joint type is a member of $\cT_n(R)$ is reproduced
correctly at the decoder, the sum of the probabilities is bounded as
\begin{eqnarray}
  \lefteqn{1-\sum_{j=1}^{N_d}e_n^{(j)}}\nonumber\\
  &\ge& \Pr\left\{\X^n\in T_{Q_{\X}}^n: Q_{\X}\in\cT_n(R)\right\} \nonumber\\
  &\ge& \sum_{Q_{\X}\in\cT^n(R)}\hspace{-4mm}(n+1)^{-|\cX^{(\cI_{N_s})}|}
        \exp\{-nD(Q_{\X}\| P_{\X})\} \nonumber\\
  &   & \label{eq:proof:lowrate:direct:1}\\
  &\ge& (n+1)^{-|\cX^{(\cI_{N_s})}|}\nonumber\\
  &   & \times\exp\left\{-n\min_{Q_{\X}\in\cT^n(R)}D(Q_{\X}\| P_{\X})\right\}\nonumber\\
  & = & \exp\left\{-n\left(\epsilon_n(1)+\min_{Q_{\X}\in\cT_n(R)}
        D(Q_{\X}\| P_{\X})\right)\right\},\nonumber
\end{eqnarray}
where Eq. (\ref{eq:proof:lowrate:direct:1}) comes from Lemma \ref{lemma:prob}.
This completes the proof of Theorem \ref{theorem:UniversalCode:lowrate:direct}.
\end{IEEEproof}

\medskip
The following converse theorem indicates that the exponent of correct decoding obtained
in Theorem \ref{theorem:UniversalCode:lowrate:direct} might not be tight.

\begin{teiri} \label{theorem:UniversalCode:lowrate:converse}
  Any FF-GCD code
  \[
    \{( \varphi_n,\wh{\varphi}_n^{(1)},\cdots,\wh{\varphi}_n^{(N_d)} )\}_{n=1}^{\infty}
  \]
  for the network $\bcS$ must satisfy
  \begin{eqnarray*}
    \lefteqn{1-\sum_{j=1}^{N_d}e_n^{(j)}
    \le \exp\Bigl[-n\Bigl\{-\epsilon_n(1)+\min_{Q_{\X}\in\cP_n(\cX^{(\cI_{N_s})})}}\\
    &   & \hspace{-3mm}\left(\left|\max_{j\in\cI_{N_d}}H(V_j|Q_j)-(R+\epsilon_n(1))
          \right|^{+}+D(Q_{\X}\| P_{\X})\right)\Biggr\}\Biggr]
  \end{eqnarray*}
  for any integer $n\ge 1$, any source $\X$ and a given coding rate $R=1/n\log M_n>0$,
  where
  \begin{eqnarray}
    \left.
    \begin{array}{l}
      Q_{\X}=Q_jV_j, \quad\forall j\in\cI_{N_d}\\
      Q_j\in\cP_n(\cX^{(\cS_j^c)}), \quad V_j\in\cV(\cX^{(\cS_j)}|Q_j)
    \end{array}
    \right\} \label{eq:theorem:lowrate:converse:1}
  \end{eqnarray}
  and $|a|^+=\max\{a,0\}$.
\end{teiri}

\begin{IEEEproof}
Note that the number of sequences to be decoded correctly for each decoder is at most
$\exp(nR)$. Here, let us consider $Q_{\X}\in\cP_n(\cX^{(\cI_{N_s})})$, $Q_j$ and $V_j$
that satisfy Eq. (\ref{eq:theorem:lowrate:converse:1}). The ratio $r_c(Q_{\X})$ of
sequences in the sequence set $T_{Q_{\X}}$ that the sequences are correctly
reproduced is at most
\begin{eqnarray}
  \lefteqn{r_c(Q_{\X})} \nonumber\\
  &\le& \min\left\{\min_{j\in\cI_{N_d}}
        \left(\frac{\exp(nR)}{|T_{V_j}^n(\x^{(\cS_j^c)})|}\right),1\right\}\nonumber\\
  &\le& \min\Bigl[\exp(nR)\cdot(n+1)^{|\cX^{(\cI_{N_s})}|}\nonumber\\
  &   & \hspace{6mm}\times\exp\{-n\max_{j\in\cI_{N_d}}H(V_j|Q_j)\},1\Bigr]
	\label{eq:proof:lowrate:converse:1}\\
  & = & \min\Bigl[\exp\{-n\{\max_{j\in\cI_{N_d}}H(V_j|Q_j)-(R+\epsilon_n(1))\}
        ,1\Bigr] \nonumber\\
  & = & \exp\left\{-n\left|\max_{j\in\cI_{N_d}}H(V_j|Q_j)
        -\left(R+\epsilon_n(1)\right)\right|^{+}\right\},\nonumber
\end{eqnarray}
where Eq. (\ref{eq:proof:lowrate:converse:1}) comes from Lemma \ref{lemma:sizeshell}.
Therefore, the probability $P_c(Q_{\X})$ such that the original sequence pair with type
$Q_{\X}$ is correctly reproduced is bounded as
\begin{eqnarray}
  \lefteqn{P_c(Q_{\X})} \nonumber\\
  &\le& r_c(Q_{\X})\Pr\{\X^n\in T_{Q_{\X}}^n\}\nonumber\\
  &\le& \exp\bigl\{-n\bigl|\max_{j\in\cI_{N_d}}H(V_j|Q_j)-\left(R+\epsilon_n(1)\right)
        \bigr|^{+}\nonumber\\
  &   & \hspace{10mm}+D(Q_{\X}\| P_{\X})\bigr\}, \label{eq:proof:lowrate:converse:2}
\end{eqnarray}
where Eq. (\ref{eq:proof:lowrate:converse:2}) comes from Lemma \ref{lemma:prob}.
Thus, the sum of the probabilities of correct decoding is obtained as
\begin{eqnarray}
  \lefteqn{1-\sum_{j=1}^{N_d}e_n^{(j)}}\nonumber\\
  &\le& \sum_{Q_{\X}\in\cP_n(\cX^{(\cI_{N_s})})}P_c(Q_{\X})\nonumber\\
  &\le& \sum_{Q_{\X}\in\cP_n(\cX^{(\cI_{N_s})})}\hspace{-6mm}\exp\bigl\{-n
        \bigl|\max_{j\in\cI_{N_d}}H(V_j|Q_j)\nonumber\\
  &   & \hspace{10mm}-\left(R+\epsilon_n(1)\right)\bigr|^{+}+D(Q_{\X}\| P_{\X})\bigr\}
	\nonumber\\
  &\le& (n+1)^{|\cX^{(\cI_{N_s})}|}\exp\bigl\{-n
	\min_{Q_{\X}\in\cP_n(\cX^{(\cI_{N_s})})}\nonumber\\
  &   & \hspace{-5mm}\Bigl(\Bigl|\max_{j\in\cI_{N_d}}H(V_j|Q_j)-\left(R+\epsilon_n(1)
	\right)\Bigr|^{+}+D(Q_{\X}\| P_{\X})\Bigr)\Bigr\},\nonumber\\
  &   & \label{eq:proof:lowrate:converse:3}\\
  & = & \exp\Bigl[-n\Bigl\{-\epsilon_n(1)+\min_{Q_{\X}\in\cP_n(\cX^{(\cI_{N_s})})}
	\nonumber\\
  &   & \hspace{-5mm}\Bigl(\Bigl|\max_{j\in\cI_{N_d}}H(V_j|Q_j)-\left(R+\epsilon_n(1)
        \right)\Bigr|^{+}+D(Q_{\X}\| P_{\X})\Bigr)\Bigr\}\Bigr]\nonumber
\end{eqnarray}
where Eq. (\ref{eq:proof:lowrate:converse:3}) comes from Lemma \ref{lemma:typecount}.
This completes the proof of Theorem \ref{theorem:UniversalCode:lowrate:converse}.
\end{IEEEproof}

\medskip
We can see that for any real value $R\ge R_f(\X|\bcS)$ and sufficiently large $n$ we have
\begin{eqnarray*}
  \lefteqn{\min_{Q_{\X}\in\cP(\cX^{(\cI_{N_s})})}
           \Bigl(\Bigl|\max_{j\in\cI_{N_d}}H(V_j|Q_j)-R\bigr|^{+}}\\
  &   & \hspace{10mm}+D(Q_{\X}\| P_{\X})\bigr)\bigr\}\\
  & = & \left|\max_{j\in\cI_{N_d}}H\left(P_{\X^{(\cS_j)}|\X^{(\cS_j^c)}}|
        P_{\X^{(\cS_j^c)}}\right)-R\right|^{+}\\
  & = & 0.
\end{eqnarray*}
On the other hand, for any real value $R<R_f(\X|\bcS)$ we have
\begin{eqnarray*}
  \lefteqn{\min_{Q_{\X}\in\cT(R)}D(Q_{\X}\| P_{\X})}\\
  &\ge& \min_{Q_{\X}\in\cP(\cX^{(\cI_{N_s})})}\bigl(\bigl|\max_{j\in\cI_{N_d}}
        H(V_j|Q_j)-R\bigr|^{+}\\
  &   & \hspace{20mm}+D(Q_{\X}\| P_{\X})\bigr)\bigr\}\\
  &\ge& 0.
\end{eqnarray*}
This implies that the exponent of correct decoding obtained in Theorem
\ref{theorem:UniversalCode:lowrate:direct} might not be tight.

\begin{hosoku}
The proof of the achievability part in the paper by Willems et al.
\cite{BroadcastSatelliteCoding} implies that any (universal) Slepian-Wolf code can be
directly utilized as a (universal) FF-GCD code. Namely, the Slepian-Wolf code is
achievable as an FF-GCD code if its coding rate satisfies $R\ge R_f(\X|\bcS)$. However,
such coding schemes cannot attain the optimal error exponent shown in Theorem
\ref{theorem:UniversalCode:converse}, since any existing construction of universal
Slepian-Wolf codes cannot attain the optimal error exponent. On the other hand, the
coding scheme presented in Section \ref{sec:construct} can attain the optimal error
exponent as shown in Theorem \ref{theorem:UniversalCode:direct}.
\end{hosoku}

%%%%%%%%
\subsection{Some special cases}
\label{sec:theorem:special}

Here, let us consider a special case where the number of decoders equals $N_d=2$.
One of the most representative examples is the (original) complementary delivery network,
where $N_s=N_d=2$, $\cS_1=\{1\}$ and $\cS_2=\{2\}$. We
have proposed a universal coding scheme for the complementary delivery network
\cite{UniversalComplementaryISIT,UniversalComplementaryPaper}, where we utilized a
bipartite graph as a codebook. The following of this subsection discusses the
relationships between the previous coding scheme and the new coding scheme shown in
Section \ref{sec:construct}.

With $N_d=2$, the coding graph $G(Q)$ can be translated into an equivalent
bipartite graph (denoted by $\wt{G}(Q)$) such that
\begin{itemize}
\item
  each vertex in one set corresponds to a sequence $\x^{(\cS_1^c)}\in T_{Q_1}^n$, and
  each vertex in the other set corresponds to a sequence
  $\x^{(\cS_2^c)}\in T_{Q_2}^n$.
\item
  each edge corresponds to a sequence set $\x^{(\cI_{N_s})}\in T_Q^n$, and the edge links
  between two vertices, each of which corresponds to the sequence subset
  $\x^{(\cS_j^c)}\in T_{Q_j}^n$ $(j=1,2)$ of the sequence set $\x^{(\cI_{N_s})}$.
\end{itemize}
\begin{figure}[t]
  \begin{center}
    \includegraphics[width=0.5\hsize]{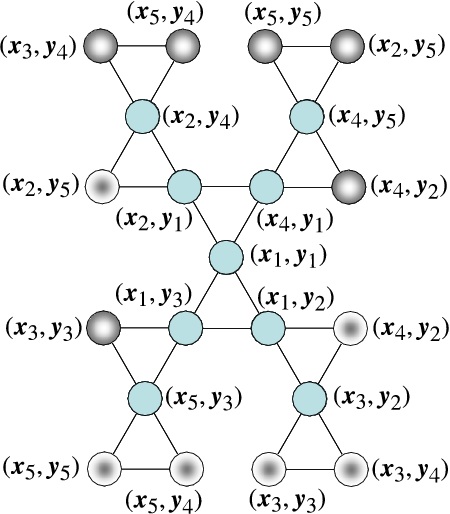}
    \caption{
      Example of the coding graph when $N_d=2$, where each vertex with a gray center
      corresponds to another vertex with a gray verge. For example, the vertex
      $(\x_3,\y_4)$ exists at the top left and the bottom right.
    }
    \label{fig:universal:CodingGraph2Decoders}
  \end{center}
\end{figure}
\begin{figure}[t]
  \begin{center}
    \includegraphics[width=0.5\hsize]{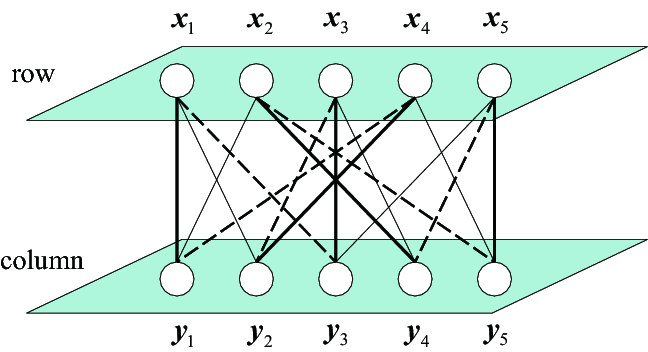}
    \caption{Bipartite graph equivalent to the coding graph shown in
             Fig. \ref{fig:universal:CodingGraph2Decoders}}
    \label{fig:universal:BipartiteGraph}
  \end{center}
\end{figure}
Fig. \ref{fig:universal:BipartiteGraph} shows an example of bipartite graphs
equivalent to the coding graph shown in Fig. \ref{fig:universal:CodingGraph2Decoders}.

From the nature of the equivalent bipartite graph $\wt{G}(Q)$, we can easily obtain
\begin{eqnarray*}
  \chi(G) &=& \chi'(\wt{G}).
\end{eqnarray*}
Therefore, the coding rate of the proposed coding scheme is determined by the edge
chromatic number $\chi'(\wt{G})$ of the equivalent bipartite graph $\wt{G}(Q)$. To this
end, we introduce the following lemmas.

\begin{hodai} \label{lemma:graph:bipartite:property}
  If the number of decoders equals $N_d=2$, then the degree of the bipartite graph
  $\wt{G}(Q)$ equivalent to the coding graph $G(Q)$ is constant for a given joint type
  $Q\in\cP_n(\cX^{(\cI_{N_s})})$, obtained as follows:
  \begin{eqnarray*}
    \Delta(\wt{G}(Q)) &=& \max_{j=1,2}|T_{V_j}^n(\x^{(\cS_j^c)})|,
  \end{eqnarray*}
  where $\x^{(\cS_j^c)}\in T_{Q_j}^n$. This equals the clique number $\omega(G(Q))$ of
  the coding graph $G(Q)$.
\end{hodai}

\begin{IEEEproof}
  We can easily obtain this lemma from the fact that the number of edges connected to the
  node $\x^{(\cS_j^c)}$ equals $|T_{V_j}^n(\x^{(\cS_j^c)})|$.
\end{IEEEproof}

\begin{hodai} \label{lemma:graph:bipartite:chromatic}
  If the number of decoders equals $N_d=2$, then for a given joint type $Q\in\cT_n(R)$
  the edge chromatic number of the bipartite graph $\wt{G}(Q)$ equivalent to the coding
  graph $G(Q)$ is bounded as
  \begin{eqnarray*}
    \chi'(\wt{G}(Q)) \le \exp(nR).
  \end{eqnarray*}
\end{hodai}

\begin{IEEEproof}
This property is directly derived from Lemmas \ref{lemma:sizeshell},
\ref{lemma:graph:konig} and \ref{lemma:graph:bipartite:property} as follows:
\begin{eqnarray}
  \chi'(\wt{G}(Q))
  & = & \Delta(\wt{G}(Q)) \label{eq:proof:bipartite:chromatic:1}\\
  & = & \max_{j=1,2} |T_{V_j}^n(\x^{(\cS_j^c)})|
        \label{eq:proof:bipartite:chromatic:2}\\
  &\le& \max_{j=1,2} \exp\{nH(V_j|Q_j)\} \label{eq:proof:bipartite:chromatic:3}\\
  &\le& \exp(nR), \label{eq:proof:bipartite:chromatic:4}
\end{eqnarray}
where Eq. (\ref{eq:proof:bipartite:chromatic:1}) comes from Lemma
\ref{lemma:graph:konig}, Eq. (\ref{eq:proof:bipartite:chromatic:2}) from Lemma
\ref{lemma:graph:bipartite:property}, Eq. (\ref{eq:proof:bipartite:chromatic:3}) from
Lemma \ref{lemma:sizeshell}, and Eq. (\ref{eq:proof:bipartite:chromatic:4}) from the
definition of $\cT_n(R)$. This concludes the proof of Lemma
\ref{lemma:graph:bipartite:chromatic}.
\end{IEEEproof}

\medskip
To summarize the above discussions, we obtain
\[
  \chi(G(Q)) = \chi'(\wt{G}(Q)) = \omega(G(Q)) \le \exp(nR).
\]

From the above discussions, we can obtain the following direct theorems for the universal
FF-GCD codes of $N_d=2$, which cannot be derived as corollaries of the theorems shown in
the previous section.

\begin{teiri} \label{theorem:UniversalCode:two:direct}
  If the number of decoders equals $N_d=2$, then for a given real number $R>0$ there
  exists a universal FF-GCD code
  \[
    \{(\varphi_n,\wh{\varphi}_{(1)}^n,\wh{\varphi}_{(2)}^n)\}_{n=1}^{\infty}
  \]
  for the network $\bcS$ such that for any integer $n\ge 1$ and any source $\X$
  \begin{eqnarray*}
    \frac 1n\log M_n &\le& R+\epsilon_n(1),\\
    e_n^{(1)}+e_n^{(2)}
    &\le& \\
    &   & \hspace{-15mm}\exp\left\{-n\left(-\epsilon_n(2)+\hspace{-2mm}
          \min_{Q_{\X}\in\cT_n^c(R)}\hspace{-2mm}D(Q_{\X}\| P_{\X})\right)\right\}.
  \end{eqnarray*}
\end{teiri}

\begin{teiri} \label{theorem:UniversalCode:two:lowrate:direct}
  For a given real number $R>0$, there exists a universal FF-GCD code
  \[
    \{(\varphi_n,\wh{\varphi}_{(1)}^n,\wh{\varphi}_{(2)}^n)\}_{n=1}^{\infty}
  \]
  for the network $\bcS$ such that for any integer $n\ge 1$ and any source $\X$ 
  \begin{eqnarray*}
    \frac 1n\log M_n &\le& R+\epsilon_n(1),\\
    1-(e_n^{(1)}+e_n^{(2)})
    &\ge& \\
    &   & \hspace{-13mm}\exp\left\{-n\left(\epsilon_n(1)+\hspace{-2mm}
          \min_{Q_{\X}\in\cT_n(R)}\hspace{-2mm}D(Q_{\X}\| P_{\X})\right)\right\},
  \end{eqnarray*}
\end{teiri}

The previous universal coding scheme for the original complementary delivery network
utilized a bipartite graph as a codebook, and derived coding theorems that were special
cases of Theorems \ref{theorem:UniversalCode:two:direct} and
\ref{theorem:UniversalCode:two:lowrate:direct}.

%%%%%%%%
\section{Variable-length coding}
\label{sec:variable}

This section discusses variable-length coding for the generalized complementary delivery
network, and shows an explicit construction of universal variable-length codes. The
coding scheme is similar to that of fixed-length codes, and also utilizes the coding
graphs defined in Section \ref{sec:construct}.

%%%%
\subsection{Formulation}
\label{sec:variable:formulate}

\begin{teigi}  \label{def:code:variable}
  {\rm(Fixed-to-variable generalized complementary delivery (FV-GCD) code)}\\
  A sequence
  \[
    \{( \varphi_n,\wh{\varphi}_n^{(1)},\cdots,\wh{\varphi}_n^{(N_d)} )\}_{n=1}^{\infty}
  \]
  of codes
  \[
    ( \varphi_n,\wh{\varphi}_n^{(1)},\cdots,\wh{\varphi}_n^{(N_d)} )
  \]
  is an FV-GCD code for the network $\bcS$ if
  \begin{eqnarray*}
    \varphi_n            &:& \cX^{(\cI_{N_s})n}\rightarrow\cB^*\\
    \wh{\varphi}_n^{(j)} &:& \varphi_n(\cX^{(\cI_{N_s})n})\times
                             \cX^{(\cS_j^c)n}\rightarrow\cX^{(\cS_j)n},
                             \quad\forall j\in\cI_{N_d},\\
    e_n^{(j)} &=& \Pr\left\{\X^{(\cS_j)n}\neq\wh{\X}^{(\cS_j)n}\right\} = 0,
                  \quad\forall j\in\cI_{N_d},
  \end{eqnarray*}
  where
  \begin{eqnarray*}
    \wh{\X}^{(\cS_j)n} &{\displaystyle\mathop{=}^{\mbox{\rm def.}}}&
      \wh{\varphi}_n^{(j)}(\varphi_n(\X^n),\X^{(\cS_j^c)n}).
  \end{eqnarray*}
  and the image of $\varphi_n$ is a prefix set.
\end{teigi}

\begin{teigi}  \label{def:rate:variable}
  {\rm(FV-GCD achievable rate)}\\
  $R$ is an FV-GCD achievable rate of the source $\X$ for the network $\bcS$ if and only
  if there exists an FV-GCD code
  \[
    \{( \varphi_n,\wh{\varphi}_n^{(1)},\cdots,\wh{\varphi}_n^{(N_d)} )\}_{n=1}^{\infty}
  \]
  for the network $\bcS$ that satisfies
  \begin{eqnarray*}
    \limsup_{n\to\infty}\frac 1n E\left[l(\varphi_n(\X^n))\right] &\le& R,
  \end{eqnarray*}
  where $l(\cdot): \cB^*\to\{1,2,3,\cdots\}$ is a length function.
\end{teigi}

\begin{teigi}  \label{def:min_rate:variable}
  {\rm(Inf FV-GCD achievable rate)}
  \begin{eqnarray*}
    \lefteqn{R_v(\X|\bcS)}\\
    &=& \inf\{R|
        R\mbox{ is an FV-GCD achievable rate of }\X\mbox{ for }\bcS\}.
  \end{eqnarray*}
\end{teigi}

%%%%
\subsection{Code construction}
\label{sec:variable:construct}

We construct universal FV-GCD codes (variable-length codes) in a similar manner to
universal FF-GCD codes (fixed-length codes). Note that the coding rate depends on the
type of sequence set to be encoded when constructing variable-length codes, whereas the
coding rate is fixed beforehand for fixed-length coding. The coding scheme is
as follows:

\noindent
[Encoding]
\begin{enumerate}
  \item
    Create a coding graph for each joint type $Q_{\X}\in\cP_n(\cX^{(\cI_{N_s})})$ and
    assign a symbol to each vertex of the coding graph $G(Q_{\X})$ in the same way as
    Steps 2 and 3 of Section \ref{sec:construct}. Note that a coding graph is created for
    every type $Q_{\X}\in\cP_n(\cX^{(\cI_{N_s})})$.
  \item
    For an input sequence set $\x^{(\cI_{N_s})}\in T_{Q_{\X}}^n$, the index assigned to
    the joint type $Q_{\X}$ is the first part of the codeword, and the symbol assigned to
    the corresponding vertex of the coding graph is determined as the second part of the
    codeword. Note that a codeword is assigned to every input sequence set
    $\x^{(\cI_{N_s})}\in\cX^{(\cI_{N_s})n}$, and the codeword length depends on the type
    of input sequence set.
\end{enumerate}

\noindent
[Decoding]\\
Decoding can be accomplished in almost the same way as the fixed-length coding. Note
that the decoder can always find the coding table used in the encoding scheme, and
therefore it can always reconstruct the original sequence.

%%%%
\subsection{Coding theorems}
\label{sec:variable:theorem}

We begin by showing a coding theorem for (non-universal) variable-length coding, which
indicates that the minimum achievable rate of variable-length coding is the same as that
of fixed-length coding.

\begin{teiri} \label{theorem:complementary:variable}
  {\rm (Coding theorem of FV-GCD code)}
  \begin{eqnarray*}
    R_v(\X|\bcS)
    &=& R_f(\X|\bcS)\\
    &=& \max_{j\in\cI_{N_d}} H(\X^{(\cS_j)}|\X^{(\cS_j^c)})
  \end{eqnarray*}
\end{teiri}

\begin{IEEEproof}
[Direct part]\\
We can apply an {\it achievable} FF-GCD code (fixed-length code) when creating an FV-GCD
code. The encoder $\varphi_n$ assigns the same codeword as that of the fixed-length code
to a sequence set $\x^{(\cI_{N_s})}\in\cX^{(\cI_{N_s})n}$ if the fixed-length code can
correctly reproduced the sequence set. Otherwise, the encoder sends the sequence set
itself as a codeword.

The above FV-GCD code can always reproduce the original sequence set at every decoder,
and it attains the desired coding rate.

\medskip
[Converse part]\\
Let an FV-GCD code
\[
  \{( \varphi_n,\wh{\varphi}_n^{(1)},\cdots,\wh{\varphi}_n^{(N_d)} )\}_{n=1}^{\infty}
\]
for the network $\bcS$ be given that satisfies the conditions of Definitions
\ref{def:code:variable} and \ref{def:rate:variable}. From Definition
\ref{def:rate:variable}, for any $\delta>0$ there exists an integer $n_1=n_1(\delta)$ and
then for all $n\ge n_1(\delta)$, we can obtain
\begin{eqnarray}
  \frac 1n E[l(\varphi_n(\X^n))] &\le& R+\delta.
  \label{eq:proof:fv:converse:1}
\end{eqnarray}
Here, let us define $A_n=\varphi_n(\X^n)$. Since the decoder $\wh{\varphi}_n^{(j)}$
$(j=1,2,\cdots,N_d)$ can always reproduce the original sequence set $\X^{(\cS_j)n}$ from
the received codeword $A_n$ and side information $\X^{(\cS_j^c)n}$, we can see that
\begin{eqnarray}
  H(\X^{(\cS_j)n}|A_n\X^{(\cS_j^c)n}) &=& 0\quad\forall j\in N_d. 
  \label{eq:proof:fv:converse:2}
\end{eqnarray}
Substituting $A_n$ into Eq.(\ref{eq:proof:fv:converse:1}), we have
\begin{eqnarray}
  n(R+\delta)
  &\ge& E[l(A_n)] \nonumber\\
  &\ge& H(A_n) \label{eq:proof:fv:converse:3}\\
  &\ge& H(A_n|\X^{(\cS_j^c)n}) \nonumber\\
  &\ge& I(\X^{(\cS_j)n};A_n|\X^{(\cS_j^c)n}) \nonumber\\
  & = & H(\X^{(\cS_j)n}|\X^{(\cS_j^c)n}), \label{eq:proof:fv:converse:4}
\end{eqnarray}
where Eq. (\ref{eq:proof:fv:converse:3}) comes from the fact that $A_n$ is a prefix set,
and Eq. (\ref{eq:proof:fv:converse:4}) from Eq. (\ref{eq:proof:fv:converse:2}).
Since we can select an arbitrarily small $\delta>0$ for a sufficient large $n$, we can
obtain
\begin{eqnarray*}
  R &\ge& \frac 1n H(\X^{(\cS_j)n}|\X^{(\cS_j^c)n})\\
    & = & H(\X^{(\cS_j)}|\X^{(\cS_j^c)}).
\end{eqnarray*}
Since the above inequality is satisfied for all $j\in\cI_{N_d}$, we obtain
\begin{eqnarray*}
  R &\ge& \max_{j\in\cI_{N_d}} H(\X^{(\cS_j)}|\X^{(\cS_j^c)}).
\end{eqnarray*}
This completes the proof of Theorem \ref{theorem:complementary:variable}.
\end{IEEEproof}

\medskip
The following direct theorem for universal coding indicates that the coding scheme
presented in the previous subsection can achieve the inf achievable rate.

\begin{teiri} \label{theorem:UniversalCode:variable:direct}
  There exists a universal FV-GCD code
  \[
    \{( \varphi_n,\wh{\varphi}_n^{(1)},\cdots,\wh{\varphi}_n^{(N_d)} )\}_{n=1}^{\infty}
  \]
  for the network $\bcS$ such that for any integer $n\ge 1$ and any source $\X$, the
  overflow probability $\ov{\rho}_n(R)$, namely the probability that codeword length per
  message sample exceeds a given real number $R>0$, is bounded as
  \begin{eqnarray*}
    \lefteqn{\ov{\rho}_n(R)}\\
    &\eqdef& \Pr\left\{l(\varphi_n(\X^n)) > nR\right\}\\
    & \le  & \exp\left\{-n\left(-\epsilon_n(N_d)+\hspace{-5mm}\min_{Q_{\X}\in\cT_n^c
             (R-\epsilon_n(N_d))}\hspace{-5mm}D(Q_{\X}\| P_{\X})\right)\right\}.
  \end{eqnarray*}
  This implies that there exists a universal FV-GCD code
  \[
    \{( \varphi_n,\wh{\varphi}_n^{(1)},\cdots,\wh{\varphi}_n^{(N_d)} )\}_{n=1}^{\infty}
  \]
  for the network $\bcS$ that satisfies
  \begin{eqnarray}
    \limsup_{n\to\infty}\frac 1n l(\varphi_n(\X^n)) &\le& R_v(\X|\bcS)
    \quad\mbox{a.s.} \label{eq:theorem:universalcode:variable:direct:1}
  \end{eqnarray}
\end{teiri}

\begin{IEEEproof}
The overflow probability can be obtained in the same way as an upperbound of the error
probability of the FF-GCD code, which has been shown in the proof of Theorem
\ref{theorem:UniversalCode:direct}. Thus, we have
\begin{eqnarray*}
  \sum_{n=1}^{\infty}\Pr\left\{\frac 1n l(\varphi_n(\X^n))>R_v(\X|\bcS)
  +\delta\right\} & < & \infty
\end{eqnarray*}
for a given $\delta>0$. From Borel-Cantelli's lemma \cite[Lemma 4.6.3]{GrayBook},
we immediately obtain Eq. (\ref{eq:theorem:universalcode:variable:direct:1}).
This completes the proof of Theorem \ref{theorem:UniversalCode:variable:direct}.
\end{IEEEproof}

\medskip\noindent
The converse theorem for variable-length coding can be easily obtained in the same way as
Theorem \ref{theorem:UniversalCode:converse}.

\begin{teiri} \label{theorem:UniversalCode:variable:converse}
  Any FV-GCD code
  \[
    \{( \varphi_n,\wh{\varphi}_n^{(1)},\cdots,\wh{\varphi}_n^{(N_d)} )\}_{n=1}^{\infty}
  \]
  for the network $\bcS$ must satisfy
  \begin{eqnarray*}
    \lefteqn{\ov{\rho}_n(R)}\\
    &\ge& \exp\left\{-n\left(\epsilon_n(2)+\min_{Q_{\X}\in\cT_n^c(R+\epsilon_n(2))}
          D(Q_{\X}\| P_{\X})\right)\right\}
  \end{eqnarray*}
  for a given real number $R>0$ and any integer $n\ge 1$.
\end{teiri}

\medskip
The following corollary is directly derived from Theorems
\ref{theorem:UniversalCode:variable:direct} and
\ref{theorem:UniversalCode:variable:converse}.

\begin{kei}
  There exists a universal FV-GCD code
  \[
    \{( \varphi_n,\wh{\varphi}_n^{(1)},\cdots,\wh{\varphi}_n^{(N_d)} )\}_{n=1}^{\infty}
  \]
  for the network $\bcS$ such that for any source $\X$
  \begin{eqnarray*}
    \limsup_{n\to\infty}\frac 1n l(\varphi_n(\X^n))
    &\le& R_v(\X|\bcS)\quad\mbox{a.s.}\\
    \lim_{n\to\infty}-\frac 1n\log\ov{\rho}_n(R)
    &=& \min_{Q_{\X}\in\cT^c(R)}D(Q_{\X}\| P_{\X})
  \end{eqnarray*}
\end{kei}

Next, we investigate the underflow probability, namely the probability that the codeword
length per message sample falls below a given real number $R>0$. For this purpose, we
present the following two theorems. The proofs are almost the same as those of Theorems
\ref{theorem:UniversalCode:lowrate:direct} and
\ref{theorem:UniversalCode:lowrate:converse}.

\begin{teiri} \label{theorem:UniversalCode:variable:under:direct}
  There exists a universal FV-GCD code
  \[
    \{( \varphi_n,\wh{\varphi}_n^{(1)},\cdots,\wh{\varphi}_n^{(N_d)} )\}_{n=1}^{\infty}
  \]
  for the network $\bcS$ such that for any integer $n\ge 1$ and any source $\X$, the
  underflow probability $\ul{\rho}_n(R)$ is bounded as
  \begin{eqnarray*}
    \lefteqn{\ul{\rho}_n(R)\eqdef\Pr\left\{l(\varphi_n(\X^n)) < nR\right\}}\\
    &\le& \\
    &   & \hspace{-13mm}\exp\left\{-n\left(\epsilon_n(1)+\min_{Q_{\X}\in
          \cT_n(R-\epsilon_n(N_d))}D(Q_{\X}\| P_{\X})\right)\right\}.
  \end{eqnarray*}
  This implies that there exists a universal FV-CD code
  \[
    \{( \varphi_n,\wh{\varphi}_n^{(1)},\cdots,\wh{\varphi}_n^{(N_d)} )\}_{n=1}^{\infty}
  \]
  for the network $\bcS$ that satisfies
  \begin{eqnarray*}
    \liminf_{n\to\infty}\frac 1n l(\varphi_n(\X^n)) &\ge& R_v(\X|\bcS)
      \quad\mbox{a.s.}
  \end{eqnarray*}
\end{teiri}

\begin{teiri} \label{theorem:UniversalCode:variable:under:converse}
  Any FV-GCD code
  \[
    \{( \varphi_n,\wh{\varphi}_n^{(1)},\cdots,\wh{\varphi}_n^{(N_d)} )\}_{n=1}^{\infty}
  \]
  for the network $\bcS$ must satisfy
  \begin{eqnarray*}
    \lefteqn{\ul{\rho}_n(R)
    \le \exp\Bigl[-n\Bigl\{-\epsilon_n(1)+\min_{Q_{\X}\in\cP_n(\cX^{(\cI_{N_s})})}}\\
    &   & \hspace{-3mm}\left(\left|\max_{j\in\cI_{N_d}}H(V_j|Q_j)-\left(R+\epsilon_n(1)
          \right)\right|^{+}+D(Q_{\X}\| P_{\X})\right)\Biggr\}\Biggr]
  \end{eqnarray*}
  for a given real number $R>0$ and any integer $n\ge 1$.
\end{teiri}

%%%%%%%%
\section{Concluding remarks}
\label{sec:conclude}

This paper dealt with a universal coding problem for a multiterminal source network
called the generalized complementary delivery network. First, we presented an explicit
construction of universal fixed-length codes, where a codebook can be expressed as a
graph and the encoding scheme is equivalent to vertex coloring of the graph. We showed
that the error exponent achieved with the proposed coding scheme is asymptotically
optimal. Next, we applied the proposed coding scheme to the construction of universal
variable-length codes. We showed that there exists a universal code such that the
codeword length converges to the minimum achievable rate almost surely.

Two important problems remains to be solved: First, the proposed coding scheme is
impractical owing to the difficulty of finding codewords from the coding table and the
substantial amount of storage space needed for the coding table. Second, this paper dealt
only with lossless coding, and therefore the construction of universal lossy codes still
remains an open problem. We have investigated the above mentioned problems for the
(original) complementary delivery network, and proposed simple coding schemes for both
lossless and lossy coding \cite{LinearUniversalComplementarySTW}. However, these coding
schemes cannot be directly extended to the generalized complementary delivery network.
Practical coding schemes for the generalized complementary delivery network should be
addressed.

%%%%%%%%
\section*{Acknowledgements}
The authors would like to thank Prof. Ryutaroh Matsumoto of Tokyo Institute of Technology
for his valuable discussions and helpful comments. The authors also thank Dr. Yoshinobu
Tonomura, Dr. Hiromi Nakaiwa, Dr. Tatsuto Takeuchi, Dr. Shoji Makino and Dr. Junji Yamato
of NTT Communication Science Laboratories for their help.

%%%%%%%%
\bibliographystyle{bib/IEEEtran}
\bibliography{bib/IEEEabrv,bib/defs,bib/it}

\begin{thebibliography}{10}
\providecommand{\url}[1]{#1}
\csname url@samestyle\endcsname
\providecommand{\newblock}{\relax}
\providecommand{\bibinfo}[2]{#2}
\providecommand{\BIBentrySTDinterwordspacing}{\spaceskip=0pt\relax}
\providecommand{\BIBentryALTinterwordstretchfactor}{4}
\providecommand{\BIBentryALTinterwordspacing}{\spaceskip=\fontdimen2\font plus
\BIBentryALTinterwordstretchfactor\fontdimen3\font minus
  \fontdimen4\font\relax}
\providecommand{\BIBforeignlanguage}[2]{{%
\expandafter\ifx\csname l@#1\endcsname\relax
\typeout{** WARNING: IEEEtran.bst: No hyphenation pattern has been}%
\typeout{** loaded for the language `#1'. Using the pattern for}%
\typeout{** the default language instead.}%
\else
\language=\csname l@#1\endcsname
\fi
#2}}
\providecommand{\BIBdecl}{\relax}
\BIBdecl

\bibitem{SlepianWolf}
D.~Slepian and J.~K. Wolf, ``Noiseless coding of correlated information
  sources,'' \emph{{IEEE} Trans. Inf. Theory}, vol.~19, no.~4, pp. 471--480,
  July 1973.

\bibitem{SideInformationCoding:Wyner}
A.~D. Wyner, ``On source coding with side information at the decoder,''
  \emph{{IEEE} Trans. Inf. Theory}, vol.~21, no.~3, pp. 294--300, May 1975.

\bibitem{KornerMarton}
J.~K\dotdot{o}rner and K.~Marton, ``Images of a set via two channels and their
  role in multi-user communication,'' \emph{{IEEE} Trans. Inf. Theory},
  vol.~23, no.~6, pp. 751--761, November 1975.

\bibitem{SgarroCoding}
A.~Sgarro, ``Source coding with side information at several decoders,''
  \emph{{IEEE} Trans. Inf. Theory}, vol.~23, no.~2, pp. 179--182, March 1977.

\bibitem{UniversalSlepianWolf:Csiszar}
I.~Csisz\dash{a}r and J.~K\dotdot{o}rner, ``Towards a general theory of source
  networks,'' \emph{{IEEE} Trans. Inf. Theory}, vol.~26, no.~2, pp. 155--165,
  March 1980.

\bibitem{UniversalSlepianWolf:Csiszar2}
I.~Csisz\dash{a}r, ``Linear codes for source and source networks: Error
  exponents, universal coding,'' \emph{{IEEE} Trans. Inf. Theory}, vol.~28,
  no.~4, pp. 585--592, July 1982.

\bibitem{UniversalSlepianWolf:Oohama}
Y.~Oohama and T.~S. Han, ``Universal coding for the {S}lepian-{W}olf data
  compression system and the strong converse theorem,'' \emph{{IEEE} Trans.
  Inf. Theory}, vol.~40, no.~6, pp. 1908--1919, November 1994.

\bibitem{UniversalSlepianWolf:Uyematsu}
T.~Uyematsu, ``An algebraic construction of codes for {S}lepian-{W}olf source
  networks,'' \emph{{IEEE} Trans. Inf. Theory}, vol.~47, no.~7, pp. 3082--3088,
  November 2001.

\bibitem{PhD:muramatsu}
J.~Muramatsu, ``Universal data compression algorithms for stationary ergodic
  sources based on the complexity of sequences,'' Ph.D. dissertation, Nagoya
  University, March 1998.

\bibitem{WynerZiv}
A.~D. Wyner and J.~Ziv, ``The rate-distortion function for source coding with
  side information at the decoder,'' \emph{{IEEE} Trans. Inf. Theory}, vol.~22,
  no.~1, pp. 1--10, January 1976.

\bibitem{BroadcastSatelliteCodingOrig}
F.~M.~J. Willems, J.~K. Wolf, and A.~D. Wyner, ``Communicating via a processing
  broadcast satellite,'' in \emph{Proc. of the 1989 IEEE/CAM Information Theory
  Workshop}, June 1989.

\bibitem{BroadcastSatelliteCoding}
A.~D. Wyner, J.~K. Wolf, and F.~M.~J. Willems, ``Communicating via a processing
  broadcast satellite,'' \emph{{IEEE} Trans. Inf. Theory}, vol.~48, no.~6, pp.
  1243--1249, June 2002.

\bibitem{NetworkCodingOrig}
R.~Ahlswede, N.~Cai, S.~R. Li, and R.~W. Yeung, ``Network information flow,''
  \emph{{IEEE} Trans. Inf. Theory}, vol.~46, no.~4, pp. 1204--1216, July 2000.

\bibitem{LinearNetworkCoding}
S.~R. Li, R.~W. Yeung, and N.~Cai, ``Linear network coding,'' \emph{{IEEE}
  Trans. Inf. Theory}, vol.~49, no.~2, pp. 371--381, February 2003.

\bibitem{NetworkSlepianWolfHan}
T.~S. Han, ``Slepian-{W}olf-{C}over theorem for networks of channels,''
  \emph{Information and Control}, vol.~47, no.~1, pp. 67--83, October 1980.

\bibitem{NetworkCodingCorrelatedBarros}
J.~Barros and S.~Servetto, ``Network information flow with correlated
  sources,'' \emph{{IEEE} Trans. Inf. Theory}, vol.~52, no.~1, pp. 155--170,
  January 2006.

\bibitem{SeparationCorrelatedNetwork}
A.~Ramamoorthy, K.~Jain, P.~A. Chou, and E.~Effros, ``Separating distributed
  source coding from network coding,'' \emph{{IEEE} Trans. Inf. Theory},
  vol.~52, no.~6, pp. 2785--2795, June 2006.

\bibitem{RandomLinearNetworkCoding}
T.~Ho, M.~M\dash{e}dard, R.~Koetter, D.~R. Karger, M.~Effros, J.~Shi, and
  B.~Leung, ``A random linear network coding approach to multicast,''
  \emph{{IEEE} Trans. Inf. Theory}, vol.~52, no.~10, pp. 4413--4430, October
  2006.

\bibitem{NetworkSlepianWolfCristescu}
R.~Cristescu, B.~Beferull-Lozaon, and M.~Vetterli, ``Networked
  {S}lepian-{W}olf: theory, algorithms, and scaling laws,'' \emph{{IEEE} Trans.
  Inf. Theory}, vol.~51, no.~12, pp. 4057--4073, December 2005.

\bibitem{LinearUniversalComplementarySTW}
S.~Kuzuoka, A.~Kimura, and T.~Uyematsu, ``Simple coding schemes for lossless
  and lossy complementary delivery problems,'' in \emph{Proc. Shannon Theory
  Workshop (STW)}, September 2007, pp. 43--50.

\bibitem{CsiszarKorner}
I.~Csisz\dash{a}r and J.~K\dotdot{o}rner, \emph{Information theory: Coding
  theorems for discrete memoryless systems}.\hskip 1em plus 0.5em minus
  0.4em\relax New York: Academic Press, 1981.

\bibitem{chromaticnumber:brooks}
R.~L. Brooks, ``On coloring the nodes of a network,'' in \emph{Proc. Cambridge
  Philos. Soc.}, vol.~37, 1941, pp. 194--197.

\bibitem{GraphTheory:biggs}
N.~L. Biggs, E.~K. Lloyd, and R.~J. Wilson, \emph{Graph {T}heory}.\hskip 1em
  plus 0.5em minus 0.4em\relax Oxford {U}niversity {P}ress, 1976.

\bibitem{chromaticnumber:vizing}
V.~G. Vizing, ``On an estimate of the chromatic class of a $p$-graph,''
  \emph{Diskret. Analiz.}, vol.~3, pp. 23--30, 1964, (in Russian).

\bibitem{EdgeColoring}
D.~K\dotdot{o}nig, ``Graphok \'{e}s alkalmaz\'{a}suk a determin\'{a}nsok \'{e}s
  a halmazok elm\'{e}let\'{e}re,'' \emph{Mathematikai \'{e}s
  Term\'{e}szettudom\'{a}nyi \'{E}rtesit\dotdot{o}}, vol.~34, pp. 104--119,
  1916, (in Hungarian).

\bibitem{UniversalComplementaryISIT}
A.~Kimura, T.~Uyematsu, and S.~Kuzuoka, ``Universal coding for correlated
  sources with complementary delivery,'' in \emph{Proc. IEEE International
  Symposium on Information Theory (ISIT)}, June 2007, pp. 1756--1760.

\bibitem{UniversalComplementaryPaper}
------, ``Universal coding for correlated sources with complementary
  delivery,'' \emph{{IEICE} Trans. Fundamentals}, vol. E90-A, no.~9, pp.
  1840--1847, September 2007.

\bibitem{GrayBook}
R.~M. Gray, \emph{Probability, {R}andom {P}rocesses, {E}rgodic
  {P}roperties}.\hskip 1em plus 0.5em minus 0.4em\relax New York:
  Springer-{V}erlag, 1988.

\end{thebibliography}
%\bibliography{bib/IEEEabrv,bib/defs,bib/test_it}

\end{document}